\documentclass[a4paper,11pt]{article}
\usepackage{pos}
\usepackage{tikz}
\usepackage{tikz-feynman}
 
 \usepackage{slashed}

\usetikzlibrary{positioning}
\usetikzlibrary{automata,positioning,arrows,matrix,backgrounds,calc}
\usetikzlibrary{decorations.text}
\usetikzlibrary{decorations.pathmorphing}
\usetikzlibrary{shapes.geometric}
\usetikzlibrary{arrows.meta}
\usepackage{pgf}
\usepackage{relsize}
\usetikzlibrary{decorations,patterns,arrows.meta}
\usetikzlibrary{decorations.pathreplacing,decorations.markings}
\tikzset{
    ->-/.style={decoration={
  markings,
  mark=at position .5 with {\arrow{>}}},postaction={decorate}},
    -<-/.style={decoration={
  markings,
  mark=at position .5 with {\arrow{<}}},postaction={decorate}},
    ->/.style={decoration={
  markings,
  mark=at position .4 with {\arrow{>}}},postaction={decorate}},
}
\definecolor{col1}{RGB}{100,143,255}
\definecolor{col2}{RGB}{120, 94, 240}
\definecolor{col3}{RGB}{254,97,0}
\definecolor{col4}{RGB}{220, 38, 127}
\definecolor{col5}{RGB}{255, 176, 0}
\usepackage{fontawesome}
\newcommand{\ep}{\varepsilon}
\newcommand{\gE}{E}
\newcommand{\gEI}{E^{\textrm{int}}}
\newcommand{\bb}{\boldsymbol}
\newcommand{\gVI}{V^{\textrm{int}}}
\newcommand{\gVE}{V^{\textrm{ext}}}
\newcommand{\gVEI}{V^{\textrm{ext}}_{\textrm{in}}}
\newcommand{\gVEO}{V^{\textrm{ext}}_{\textrm{out}}}
\newcommand{\gGE}{{\Gamma^{\textrm{ext}}}}
\newcommand{\gGI}{{\Gamma^{\textrm{int}}}}

\newcommand{\sunny}{\text{\faSunO}}
\newcommand{\shady}{\text{\faMoonO}}

\DeclareMathOperator{\sinc}{sinc}
  
\newcommand{\dd}{\textrm{d}}
\newcommand{\pp}{\textrm{p}}
\renewcommand{\tt}{\textrm{t}}
\newcommand{\cc}{\textrm{c}}
\newcommand{\rr}{\textrm{r}}
\DeclareMathOperator{\Sym}{Sym}
\title{Flow Oriented Perturbation Theory}
\author*[a]{Alexandre Salas-Bernárdez}
\author[b]{Michael Borinsky}
\author[c]{Zeno Capatti}
\author[4,5,d]{Eric Laenen}

\affiliation[a]{Universidad Complutense de Madrid, Departamento de F\'isica Te\'orica and IPARCOS, 28040 Madrid, Spain.}

\affiliation[c]{Institute for Theoretical Physics, ETH Z\"urich, 8093 Z\"urich, Switzerland.}

\affiliation[b]{Institute for Theoretical Studies, ETH Z\"urich, 8092 Z\"urich, Switzerland.}

\affiliation[4]{Nikhef, Theory Group, Science Park 105, 1098 XG, Amsterdam, The Netherlands.}

\affiliation[3]{IOP/ITFA, University of Amsterdam, Science Park 904, 1098 XH Amsterdam, The Netherlands.}

\affiliation[d]{ITF, Utrecht University, Leuvenlaan 4, 3584 CE Utrecht, The Netherlands.}

\emailAdd{alexsala@ucm.es}

\abstract{Flow Oriented Perturbation Theory (FOPT) is a novel approach to Feynman diagrams based on the coordinate (position) space description of Quantum Field Theories (QFT). FOPT offers interesting features regarding the computation of higher-loop Feynman amplitudes such as combinatorial and canonical Feynman rules, explicit infrared singularity factorization on a per-diagram level and the potential to have manifest cancellation of real and virtual singularities. In these proceedings we briefly summarize the derivation of FOPT and present its Feynman rules for covariant diagrams, S-matrix elements and cut diagrams in massless scalar QFT, supported by examples. We then discuss the extension of FOPT to massless fermion fields and indicate steps towards the treatment of massive lines in arbitrary dimensions.}
 
\FullConference{16th International Symposium on Radiative Corrections: Applications of Quantum Field Theory to Phenomenology (
RADCOR2023)\\
28th May - 2nd June, 2023\\
Crieff, Scotland, UK\\}

\begin{document}
\maketitle

\section{FOPT's Feynman Rules for massless scalar QFT}

\subsection{Feynman rules for coordinate space amplitudes}
\label{sec:flowfeynmanrules}
Flow Oriented Perturbation Theory \cite{Borinsky:2022msp} provides an 
alternative perturbative decomposition of correlation functions as
\begin{align}
\label{eq:FOPT_greens_functions}
\Gamma(x_1,...,x_{|\gVE|})=\left\langle 0 | T ( \varphi(x_1) \cdots \varphi(x_{|\gVE|}) ) | 0 \right \rangle =
\sum_{(G,\bb \sigma)} \frac{1}{\Sym (G,\bb \sigma)} A_{G,\bb \sigma}(x_1, \ldots, x_{|\gVE|})\;,
\end{align}
where the sum runs over all topologically different directed graphs (digraphs), $(G,\bb \sigma)$, \textit{i.e.}~graphs $G$ with a specified energy flow on each propagator (an orientation $\bb \sigma$). This representation is obtained by performing all time integrations over internal vertices, $\int dy_v^0$, of a given Feynman integral corresponding to graph $G$ with internal (external)  vertices $\gVI$ ($\gVE$) and edges $\gE$, 
\begin{align}
A_G(x_1, \ldots, x_{|\gVE|})=
\frac{(-ig)^{|\gVI|}}{(2\pi)^{2|\gE|}}
\left[ \prod_{v \in \gVI} \int \dd^4 y_v \right]
\prod_{e \in \gE}\frac{1}{-z_e^2+i\eta}\;.\label{eq:FI}
\end{align}
In performing the time integrations, an individual covariant Feynman integral will be expressed into its different energy flow-oriented components:
\begin{align}
\frac{1}{\Sym G} A_{G}(x_1, \ldots, x_{|\gVE|})
=
\sum_{\langle \bb \sigma \rangle}
\frac{1}{\Sym (G,\bb \sigma)} A_{G,\bb \sigma}(x_1, \ldots, x_{|\gVE|})\;.
\end{align}

The integral expression for $A_{G,\bb \sigma}(x_1, \ldots, x_{|\gVE|})$ can be found using the following Feynman rules \cite{Borinsky:2022msp}:
\begin{enumerate}
        \item $A_{G,\bb \sigma} =0$ if the digraph $(G,\bb \sigma)$ is not energy-conserving, \textit{i.e.} the completed digraph $(G,\bb \sigma)^\circ$ (found by joining all external vertices in the special vertex $\circ$) is not strongly connected. 
        \item Multiply by a factor of $-i g$ for each interaction vertex. 
        \item For each edge $e$ of $G$ multiply by a factor $\frac{-i}{(8\pi^2)|\vec{z}_e|}$ where $\vec{z}_e = \vec{y}_{v} - \vec{y}_{u}$ and $\vec{y}_v,\vec{y}_u$ are the coordinates of the internal or external vertices to which the edge $e$ is incident. 
        \item For each admissible energy-flow path, $\pp$, of $(G,\bb \sigma)$ (i.e.~for each energy cycle in the 
        canonical cycle basis of $(G,\bb \sigma)^\circ$) multiply by a factor of $
i/\left(
{
\gamma_\pp
+
\tau_{\pp} + i\eta
}
\right),
$ where
\begin{equation}
    \gamma_\mathrm{p} =\sum_{e\in \pp}
|\vec{z}_e|
\end{equation} is the sum over all edge lengths that are in the path $\pp$ and $\tau_{\pp}$ is either the time passed between the starting and ending external vertices of the path or vanishes if the cycle does not go through the $\circ$ vertex.
        \item For each internal vertex $v$ of the graph $G$ integrate over three-dimensional space $\int \dd^3 \vec{y}_v$ and multiply by $2 \pi$.
%    \end{enumerate}
\end{enumerate}

We can summarize these Feynman rules as follows. For a given 
digraph $(G,\bb \sigma)$ with cycle basis $\Gamma$, where all interaction vertices in $G$ are internal vertices and vice-versa, we have
\begin{equation}
    A_{G,\bb \sigma}(x_1, \ldots, x_{|\gVE|})=
\frac{(2\pi g)^{|\gVI|}}{(-4\pi^2)^{|\gE|}}
\left( \prod_{v \in \gVI} \int \dd^3 \vec y_v 
\right)
\left(
\prod_{e\in \gE}
\frac{1}{2 |\vec z_e |}
\right)
 \prod_{\pp \in \Gamma} 
\frac{1}
{
\gamma_\pp
+\tau_{\pp} 
+ i \eta
}\;.
\label{eq:delta_free_rep}
\end{equation}
We next illustrate the application  of these rules in a specific example.

\paragraph{Triangle example} To illustrate the FOPT Feynman rules for covariant diagrams, we will consider the following covariant graph
%%%%%%---------------------------------------------%%%%%%%%%%%%%%%%%%%%%%%%%%%%%%%%%%%%%%%%%%%%%%%%%%%%%%%%%%%%%%%%%%%%%%%%%%%%%%%%%%%%%%%%%%%%%%%%%%%%%%%%%%%%%%%%%%%%%%%%%%%
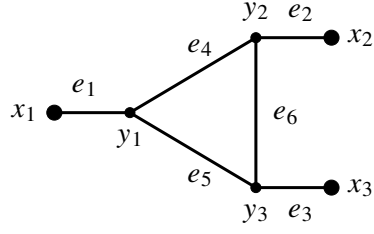
\begin{equation}
\resizebox{5cm}{!}{%
\begin{tikzpicture}
    
    %\node[] (sing) [below left = 2.72cm and 0.5cm of 1] {$ \scalebox{1.3}{\boldsymbol{\otimes}}$};
    %\node[] (3) [below right = 3cm and 3cm of 1] {};
    %\node[] (4) [below left = 3cm and 3cm of 1] {};

    \begin{feynman}
        \vertex(1);
        \vertex[above right = 1.5cm and 2.5cm of 1](2);
        \vertex[below right = 1.5cm and 2.5cm of 1](3);
        
        \vertex[left = 1.5cm of 1](E1);
        \vertex[right = 1.5cm of 2](E2);
        \vertex[right = 1.5cm of 3](E3);
        
        \vertex[left = 0.2cm of E1](L1) {\scalebox{1.5}{$x_1$}};
        \vertex[right = 0.2cm of E2](L2) {\scalebox{1.5}{$x_2$}};
        \vertex[right = 0.2cm of E3](L3) {\scalebox{1.5}{$x_3$}};
        
        \vertex[below = 0.2cm of 1](L1_I) {\scalebox{1.5}{$y_1$}};
        \vertex[above = 0.2cm of 2](L2_I) {\scalebox{1.5}{$y_2$}};
        \vertex[below = 0.2cm of 3](L3_i) {\scalebox{1.5}{$y_3$}};
        
        \vertex[above left = 0.2cm and 0.5cm of 1](edge1) {\scalebox{1.5}{$e_1$}};
        \vertex[above right = 0.2cm and 0.5cm of 2](edge2) {\scalebox{1.5}{$e_2$}};
        \vertex[below right = 0.2cm and 0.5cm of 3](edge3) {\scalebox{1.5}{$e_3$}};
        
        \vertex[above right = 1cm and 1cm of 1](edge4) {\scalebox{1.5}{$e_4$}};
        \vertex[below right = 1cm and 1cm of 1](edge5) {\scalebox{1.5}{$e_5$}};
        \vertex[below right = 1.25cm and 0.2cm of 2](edge6) {\scalebox{1.5}{$e_6$}};

         \diagram*[large]{	
            (1) -- [-,line width=0.6mm] (2) -- [-,line width=0.6mm] (3) -- [-,line width=0.6mm] (1),

            (1) -- [-,line width=0.6mm] (E1),
            (2) -- [-,line width=0.6mm] (E2),
            (3) -- [-,line width=0.6mm] (E3),

        }; 
       \path[draw=black, fill=black] (E1) circle[radius=0.15];
       \path[draw=black, fill=black] (E2) circle[radius=0.15];
       \path[draw=black, fill=black] (E3) circle[radius=0.15];
       \path[draw=black, fill=black] (1) circle[radius=0.1];
       \path[draw=black, fill=black] (2) circle[radius=0.1];
       \path[draw=black, fill=black] (3) circle[radius=0.1];
       
    \end{feynman}

    	%\draw[->, thick, black] (A3) to[out= -10, in=-170] (3);
\end{tikzpicture}
}\nonumber
\end{equation}
For the case where energy flows into the diagram through edge $e_1$, this graph has 12 energy-conserving orientations and 6 distinct configurations under the simultaneous replacement of $x_2\leftrightarrow x_3$ and $y_2\leftrightarrow y_3$. These are:
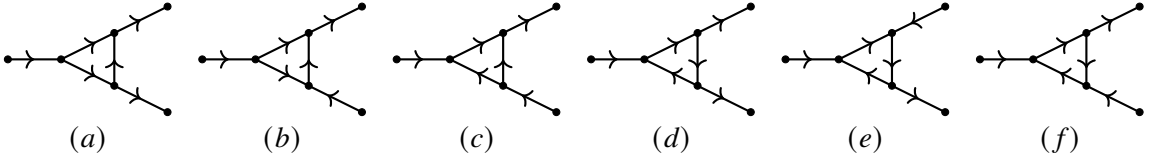
\begin{align}
\centering\nonumber
\begin{array}{cccccccccccc}
\def\scale{.7}
\begin{tikzpicture}[baseline={([yshift=-0.7ex]0,0)}] 
\coordinate (i111) at (-2*\scale,0);
    \coordinate (i1) at (-\scale,0);
    \coordinate[] (i2) at (0,.5*\scale);
    \coordinate[] (i21) at (0,-.5*\scale);
    \coordinate[] (v1) at (\scale,\scale);
    \coordinate[] (v2) at (\scale,-\scale);
    \draw[->-,thick] (i111) -- (i1) node[midway,above left] {};
    \draw[-<-,thick] (i2) -- (i1) node[midway,above left] {};
    \draw[-<-,thick] (i21) -- (i1) node[midway,below left] {};
    \draw[->-,thick] (i21) -- (v2) node[midway,below left] {};
    \draw[->-,thick] (i21) -- (i2) node[midway,right] {};
    \draw[->-,thick] (i2) -- (v1) node[midway,above left] {};
        \filldraw (v1) circle (1.3pt);
        \filldraw (i21) circle (1.3pt);
    \filldraw (v2) circle (1.3pt);
    \filldraw (i1) circle (1.3pt);
    \filldraw (i2) circle (1.3pt);
     \filldraw (i111) circle (1.3pt);
\end{tikzpicture} &\def\scale{.7}
\begin{tikzpicture}[baseline={([yshift=-0.7ex]0,0)}] 
\coordinate (i111) at (-2*\scale,0);
    \coordinate (i1) at (-\scale,0);
    \coordinate[] (i2) at (0,.5*\scale);
    \coordinate[] (i21) at (0,-.5*\scale);
    \coordinate[] (v1) at (\scale,\scale);
    \coordinate[] (v2) at (\scale,-\scale);
    \draw[->-,thick] (i111) -- (i1) node[midway,above left] {};
 \draw[-<-,thick] (i2) -- (i1) node[midway,above left] {};
    \draw[-<-,thick] (i21) -- (i1) node[midway,below left] {};
    \draw[-<-,thick] (i21) -- (v2) node[midway,below left] {};
    \draw[->-,thick] (i21) -- (i2) node[midway,right] {};
    \draw[->-,thick] (i2) -- (v1) node[midway,above left] {};
        \filldraw (v1) circle (1.3pt);
        \filldraw (i21) circle (1.3pt);
    \filldraw (v2) circle (1.3pt);
    \filldraw (i1) circle (1.3pt);
    \filldraw (i2) circle (1.3pt);
     \filldraw (i111) circle (1.3pt);
\end{tikzpicture} & \def\scale{.7}
\begin{tikzpicture}[baseline={([yshift=-0.7ex]0,0)}] 
\coordinate (i111) at (-2*\scale,0);
    \coordinate (i1) at (-\scale,0);
    \coordinate[] (i2) at (0,.5*\scale);
    \coordinate[] (i21) at (0,-.5*\scale);
    \coordinate[] (v1) at (\scale,\scale);
    \coordinate[] (v2) at (\scale,-\scale);
    \draw[->-,thick] (i111) -- (i1) node[midway,above left] {};
 \draw[-<-,thick] (i2) -- (i1) node[midway,above left] {};
    \draw[->-,thick] (i21) -- (i1) node[midway,below left] {};
    \draw[-<-,thick] (i21) -- (v2) node[midway,below left] {};
    \draw[->-,thick] (i21) -- (i2) node[midway,right] {};
    \draw[->-,thick] (i2) -- (v1) node[midway,above left] {};
        \filldraw (v1) circle (1.3pt);
        \filldraw (i21) circle (1.3pt);
    \filldraw (v2) circle (1.3pt);
    \filldraw (i1) circle (1.3pt);
    \filldraw (i2) circle (1.3pt);
     \filldraw (i111) circle (1.3pt);
\end{tikzpicture}& \def\scale{.7}
\begin{tikzpicture}[baseline={([yshift=-0.7ex]0,0)}] 
\coordinate (i111) at (-2*\scale,0);
    \coordinate (i1) at (-\scale,0);
    \coordinate[] (i2) at (0,.5*\scale);
    \coordinate[] (i21) at (0,-.5*\scale);
    \coordinate[] (v1) at (\scale,\scale);
    \coordinate[] (v2) at (\scale,-\scale);
    \draw[->-,thick] (i111) -- (i1) node[midway,above left] {};
 \draw[-<-,thick] (i2) -- (i1) node[midway,above left] {};
    \draw[->-,thick] (i21) -- (i1) node[midway,below left] {};
    \draw[->-,thick] (i21) -- (v2) node[midway,below left] {};
    \draw[-<-,thick] (i21) -- (i2) node[midway,right] {};
    \draw[->-,thick] (i2) -- (v1) node[midway,above left] {};
        \filldraw (v1) circle (1.3pt);
        \filldraw (i21) circle (1.3pt);
    \filldraw (v2) circle (1.3pt);
    \filldraw (i1) circle (1.3pt);
    \filldraw (i2) circle (1.3pt);
     \filldraw (i111) circle (1.3pt);
\end{tikzpicture}&\def\scale{.7}
\begin{tikzpicture}[baseline={([yshift=-0.7ex]0,0)}] 
\coordinate (i111) at (-2*\scale,0);
    \coordinate (i1) at (-\scale,0);
    \coordinate[] (i2) at (0,.5*\scale);
    \coordinate[] (i21) at (0,-.5*\scale);
    \coordinate[] (v1) at (\scale,\scale);
    \coordinate[] (v2) at (\scale,-\scale);
    \draw[->-,thick] (i111) -- (i1) node[midway,above left] {};
    \draw[-<-,thick] (i2) -- (i1) node[midway,above left] {};
    \draw[->-,thick] (i21) -- (i1) node[midway,below left] {};
    \draw[->-,thick] (i21) -- (v2) node[midway,below left] {};
    \draw[-<-,thick] (i21) -- (i2) node[midway,right] {};
    \draw[-<-,thick] (i2) -- (v1) node[midway,above left] {};
        \filldraw (v1) circle (1.3pt);
        \filldraw (i21) circle (1.3pt);
    \filldraw (v2) circle (1.3pt);
    \filldraw (i1) circle (1.3pt);
    \filldraw (i2) circle (1.3pt);
     \filldraw (i111) circle (1.3pt);
\end{tikzpicture} &\def\scale{.7}
\begin{tikzpicture}[baseline={([yshift=-0.7ex]0,0)}] 
\coordinate (i111) at (-2*\scale,0);
    \coordinate (i1) at (-\scale,0);
    \coordinate[] (i2) at (0,.5*\scale);
    \coordinate[] (i21) at (0,-.5*\scale);
    \coordinate[] (v1) at (\scale,\scale);
    \coordinate[] (v2) at (\scale,-\scale);
     \draw[->-,thick] (i111) -- (i1) node[midway,above left] {};
    \draw[-<-,thick] (i2) -- (i1) node[midway,above left] {};
    \draw[->-,thick] (i21) -- (i1) node[midway,below left] {};
    \draw[-<-,thick] (i21) -- (v2) node[midway,below left] {};
    \draw[-<-,thick] (i21) -- (i2) node[midway,right] {};
    \draw[->-,thick] (i2) -- (v1) node[midway,above left] {};
        \filldraw (v1) circle (1.3pt);
        \filldraw (i21) circle (1.3pt);
    \filldraw (v2) circle (1.3pt);
    \filldraw (i1) circle (1.3pt);
    \filldraw (i111) circle (1.3pt);
    \filldraw (i2) circle (1.3pt);
\end{tikzpicture}\\
(a)&(b)&(c)&(d)&(e)&(f)
\end{array}
\end{align}
The first orientation $(a)$ is decomposed into its canonical cycle basis, $\{\pp_1,\pp_2,\pp_3
\}$,  as
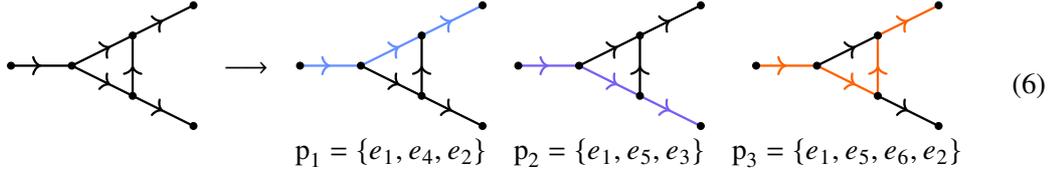
\begin{align}
\label{eq:triangle_paths}
\centering
\begin{array}{cccccccccccc}
&\def\scale{.8}
\begin{tikzpicture}[baseline={([yshift=-0.7ex]0,0)}] 
\coordinate (i00) at (-2*\scale,0);
    \coordinate (i1) at (-\scale,0);
\draw[->-,line width=0.3mm] (i00) -- (i1) node[midway,right] {};
    \filldraw (i00) circle (1.3pt);
    \coordinate[] (i2) at (0,.5*\scale);
    \coordinate[] (i21) at (0,-.5*\scale);
    \coordinate[] (v1) at (\scale,\scale);
    \coordinate[] (v2) at (\scale,-\scale);
    \draw[-<-,line width=0.3mm] (i2) -- (i1) node[midway,above left] {};
    \draw[-<-,line width=0.3mm] (i21) -- (i1) node[midway,below left] {};
    \draw[->-,line width=0.3mm] (i21) -- (v2) node[midway,below left] {};
    \draw[->-,line width=0.3mm] (i21) -- (i2) node[midway,right] {};
    \draw[->-,line width=0.3mm] (i2) -- (v1) node[midway,above left] {};
        \filldraw (v1) circle (1.3pt);
        \filldraw (i21) circle (1.3pt);
    \filldraw (v2) circle (1.3pt);
    \filldraw (i1) circle (1.3pt);
    \filldraw (i2) circle (1.3pt);
\end{tikzpicture}&\longrightarrow&\def\scale{.8}
\begin{tikzpicture}[baseline={([yshift=-0.7ex]0,0)}] 
\coordinate (i00) at (-2*\scale,0);
    \coordinate (i1) at (-\scale,0);
\draw[col1,->-,line width=0.3mm] (i00) -- (i1) node[midway,right] {};
    \filldraw (i00) circle (1.3pt);
    \coordinate[] (i2) at (0,.5*\scale);
    \coordinate[] (i21) at (0,-.5*\scale);
    \coordinate[] (v1) at (\scale,\scale);
    \coordinate[] (v2) at (\scale,-\scale);
    \draw[col1,-<-,line width=0.3mm] (i2) -- (i1) node[midway,above left] {};
    \draw[-<-,line width=0.3mm] (i21) -- (i1) node[midway,below left] {};
    \draw[->-,line width=0.3mm] (i21) -- (v2) node[midway,below left] {};
    \draw[->-,line width=0.3mm] (i21) -- (i2) node[midway,right] {};
    \draw[col1,->-,line width=0.3mm] (i2) -- (v1) node[midway,above left] {};
        \filldraw (v1) circle (1.3pt);
        \filldraw (i21) circle (1.3pt);
    \filldraw (v2) circle (1.3pt);
    \filldraw (i1) circle (1.3pt);
    \filldraw (i2) circle (1.3pt);
\end{tikzpicture}&\def\scale{.8}
\begin{tikzpicture}[baseline={([yshift=-0.7ex]0,0)}] 
\coordinate (i00) at (-2*\scale,0);
    \coordinate (i1) at (-\scale,0);
\draw[col2,->-,line width=0.3mm] (i00) -- (i1) node[midway,right] {};
    \filldraw (i00) circle (1.3pt);
    \coordinate[] (i2) at (0,.5*\scale);
    \coordinate[] (i21) at (0,-.5*\scale);
    \coordinate[] (v1) at (\scale,\scale);
    \coordinate[] (v2) at (\scale,-\scale);
    \draw[-<-,line width=0.3mm] (i2) -- (i1) node[midway,above left] {};
    \draw[col2,-<-,line width=0.3mm] (i21) -- (i1) node[midway,below left] {};
    \draw[col2,->-,line width=0.3mm] (i21) -- (v2) node[midway,below left] {};
    \draw[->-,line width=0.3mm] (i21) -- (i2) node[midway,right] {};
    \draw[->-,line width=0.3mm] (i2) -- (v1) node[midway,above left] {};
        \filldraw (v1) circle (1.3pt);
        \filldraw (i21) circle (1.3pt);
   \filldraw (v2) circle (1.3pt);
    \filldraw (i1) circle (1.3pt);
    \filldraw (i2) circle (1.3pt);
\end{tikzpicture}&\def\scale{.8}
\begin{tikzpicture}[baseline={([yshift=-0.7ex]0,0)}] 
\coordinate (i00) at (-2*\scale,0);
    \coordinate (i1) at (-\scale,0);
\draw[col3,->-,line width=0.3mm] (i00) -- (i1) node[midway,right] {};
    \filldraw (i00) circle (1.3pt);
    \coordinate[] (i2) at (0,.5*\scale);
    \coordinate[] (i21) at (0,-.5*\scale);
    \coordinate[] (v1) at (\scale,\scale);
    \coordinate[] (v2) at (\scale,-\scale);
    \draw[-<-,line width=0.3mm] (i2) -- (i1) node[midway,above left] {};
    \draw[col3,-<-,line width=0.3mm] (i21) -- (i1) node[midway,below left] {};
    \draw[->-,line width=0.3mm] (i21) -- (v2) node[midway,below left] {};
    \draw[col3,->-,line width=0.3mm] (i21) -- (i2) node[midway,right] {};
    \draw[col3,->-,line width=0.3mm] (i2) -- (v1) node[midway,above left] {};
        \filldraw (v1) circle (1.3pt);
        \filldraw (i21) circle (1.3pt);
    \filldraw (v2) circle (1.3pt);
    \filldraw (i1) circle (1.3pt);
    \filldraw (i2) circle (1.3pt);
\end{tikzpicture}\\
&&&\pp_1=\{e_1,e_4,e_2\}&\pp_2=\{e_1,e_5,e_3\}&\pp_3=\{e_1,e_5,e_6,e_2\}
\end{array}
\end{align}
Using the energy-flow-oriented Feynman rules we obtain that this orientation equals
\begin{align}
A_{{\bb{\sigma}}_{(1)}}(x_1,x_2,x_3)=&\frac{g^3}{(2\pi)^9}\int d^3{\vec{y}}_1d^3{\vec{y}}_2 d^3{\vec{y}}_3\left(\prod_{i=1}^5\frac{1}{2|{\vec{z}}_i |}\right)\frac{1}{|{\vec{ z}}_1|+|{\vec{ z}}_4|+|{\vec{ z}}_2|+x_2^0-x_1^0+i\eta}\times\nonumber\\
&\times\frac{1}{|{\vec{z}}_1|+|{\vec{z}}_5|+|{\vec{z}}_3|+x_3^0-x_1^0+i\eta}\,\frac{1}{|{\vec{z}}_1|+|{\vec{z}}_5|+|{\vec{ z}}_6|+|{\vec{ z}}_2|+x_2^0-x_1^0+i\eta}\;.\label{eq:firstorientationtriangle}
\end{align}

\subsection{FOPT representation of S-matrix elements}

\label{sec:px_derivation}

The FOPT representation of the S-matrix follows by applying a similar treatment of Feynman graphs to S-matrix elements \cite{Borinsky:2022msp}, so that an S-matrix element can be expressed as follows
\begin{equation}
    S(\{p_i\}_{i\in \gVEI}, \{p_f\}_{f\in \gVEO})=
\sum_{(G,\bb \sigma)} \frac{1}{\Sym (G,\bb \sigma)} 
    S_{G,\bb \sigma}(\{p_i\}_{i\in \gVEI}, \{p_f\}_{f\in \gVEO})\;,
\end{equation}
where we sum over all FOPT graphs (energy orientations). In \cite{Borinsky:2022msp} we regard this representation as the $p$-$x$ representation of the S-matrix, since the external kinematics are given in momentum space, whereas internal integrations are performed in coordinate space.

Picking a reference internal vertex $w \in \gVI$ of the graph. An S-matrix element for a given orientation of a graph $G$ equals,
\begin{align}
\label{eq:x-p_representation}
S_{G,\bb \sigma}
=
\frac{Z^{|V^{\text{ext}}|/2}
(2\pi)^3
(2\pi g)^{|\gVI|}
i^{|\gVE|}
}{(-4\pi^2)^{|\gE|}i^{|\gGE|}}
\delta^{(4)}\left(
\sum_{a \in \gVE} p_a
\right)
s_{G,\bb \sigma}(\{p_i\}_{i\in \gVEI}, \{p_f\}_{f\in \gVEO})
\;,
\end{align}
where $Z$ are renormalization constants, and 
 $
s_{G,\bb \sigma}=(\{p_i\}_{i\in \gVEI}, \{p_f\}_{f\in \gVEO})
$ is the \emph{reduced} S-matrix element without trivial prefactors, 
\begin{gather}
\begin{gathered}
\label{eq:sG_red}
s_{G,\bb \sigma}
=
\int
\left[
\frac{
\prod_{v\in V^{\text{int}}\setminus\{w\}} \mathrm{d}^3 \vec{y}_v
}{
\prod_{e\in \gEI} 2|\vec{z}_e|
}
\right]
\left[
  \frac{
\prod_{a\in \gVE} 
e^{-i\vec{y}_{\,\overline{a}} \cdot \vec{p}_a}
 }{
\left[ \prod_{\cc\in \gGI} \gamma_\cc \right]
}
\widehat{\mathcal{F}}_{G,\bb \sigma}^{\{p^0_a\}}(\bb \gamma^\tt + i \ep \bb 1)
\right]
\Bigg|_{\vec y_w = 0}.
\end{gathered}
\end{gather}
In eq. (\ref{eq:sG_red}), $\gamma_\cc$ are the path lengths corresponding to paths that do not have external edges (which we regard as \textit{cycles}, $\gGI$), $\vec{y}_{\,\overline{a}}$ are the internal vertices adjacent to vertex $a$, and $\widehat{\mathcal{F}}_{G,\bb \sigma}^{\{p^0_a\}}(\bb \gamma^\tt + i \ep \bb 1)$ is the Fourier transform of the so called \textit{flow polytope}, $\mathcal{F}_{G,\bb \sigma}^{\{p^0_a\}}$. Here $\gamma^\tt $ are the path lengths of truncated routes $\rr^\tt$, \textit{i.e.} paths that do have external edges but with their length subtracted. We will elucidate with the next example how to construct the flow polytope of a given orientation.

\paragraph{Triangle example}

To illustrate the $p$-$x$ representation of the S-matrix we discuss here the contribution to the S-matrix of the orientation $(a)$ in the triangle example above,
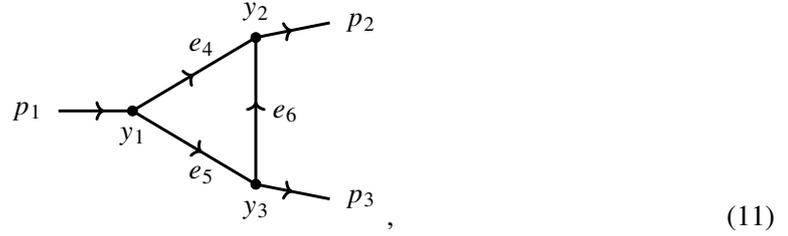
\begin{equation}
\label{eq:Striangle}
\resizebox{5cm}{!}{%
\begin{tikzpicture}
    
    %\node[] (sing) [below left = 2.72cm and 0.5cm of 1] {$ \scalebox{1.3}{\boldsymbol{\otimes}}$};
    %\node[] (3) [below right = 3cm and 3cm of 1] {};
    %\node[] (4) [below left = 3cm and 3cm of 1] {};

    \begin{feynman}
        \vertex(1);
        \vertex[above right = 1.5cm and 2.5cm of 1](2);
        \vertex[below right = 1.5cm and 2.5cm of 1](3);
        
        \vertex[left = 1.5cm of 1](E1);
        \vertex[above right = 0.3cm and 1.5cm of 2](E2);
        \vertex[below right = 0.3cm and 1.5cm of 3](E3);
        
        \vertex[left = 0.2cm of E1](L1) {\scalebox{1.5}{$p_1$}};
        \vertex[right = 0.2cm of E2](L2) {\scalebox{1.5}{$p_2$}};
        \vertex[right = 0.2cm of E3](L3) {\scalebox{1.5}{$p_3$}};
        
        \vertex[below = 0.2cm of 1](L1_I) {\scalebox{1.5}{$y_1$}};
        \vertex[above = 0.2cm of 2](L2_I) {\scalebox{1.5}{$y_2$}};
        \vertex[below = 0.2cm of 3](L3_i) {\scalebox{1.5}{$y_3$}};
        
        \vertex[above right = 1cm and 1cm of 1](edge4) {\scalebox{1.5}{$e_4$}};
        \vertex[below right = 1cm and 1cm of 1](edge5) {\scalebox{1.5}{$e_5$}};
        \vertex[below right = 1.25cm and 0.2cm of 2](edge6) {\scalebox{1.5}{$e_6$}};

         \diagram*[large]{	
            (1) -- [->-,line width=0.6mm] (2) -- [-<-,line width=0.6mm] (3) -- [-<-,line width=0.6mm] (1),

            (1) -- [-<-,line width=0.6mm] (E1),
            (2) -- [->-,line width=0.6mm] (E2),
            (3) -- [->-,line width=0.6mm] (E3),

        }; 
       \path[draw=black, fill=black] (1) circle[radius=0.1];
       \path[draw=black, fill=black] (2) circle[radius=0.1];
       \path[draw=black, fill=black] (3) circle[radius=0.1];
       
    \end{feynman}

    	%\draw[->, thick, black] (A3) to[out= -10, in=-170] (3);
\end{tikzpicture}
},
\end{equation}
where we now label the external vertex $x_i$ with its Fourier conjugate momentum $p_i$. Our convention is that we take $p_1^0 > 0$ and $p_2^0, p_3^0<0$. The routes of this digraph have been illustrated in eq.~\eqref{eq:triangle_paths}. The three corresponding truncated routes are $\rr_1^\tt = \{e_4\}$, $\rr_2^\tt=\{e_5\}$ and $\rr_3^\tt=\{e_5,e_6\}$. Hence, $\gamma^\tt_1 = |\vec z_4|, \gamma^\tt_2 = |\vec z_5|, \gamma^\tt_3 = |\vec z_5| + |\vec z_6|$. Let $E_1, E_2$ and $E_3$ be the energies that flow through the respective route.

The flow polytope $\mathcal{F}_{G,\bb \sigma}^{\{p^0_a\}}$ for this digraph is defined by the conditions,
\begin{align}
\label{eq:path_energy_cons}
&E_1, E_2, E_3 \geq 0\,,&E_1 + E_2 + E_3 = p_1^0\,,&&E_1 + E_3 = -p_2^0\,,&&E_2 = -p_3^0\,,&
\end{align}
where one of the last three equations is redundant by overall momentum conservation. We can give an interpretation to the energy-conservation condition of eq.~\eqref{eq:path_energy_cons} as follows: for each external vertex $v\in V^{\text{ext}}$, enumerate the paths that start or end at that vertex, and correspondingly sum their energies. Then, set the sum of such energies to be $p_v^0$ if the vertex is the starting vertex for such paths or $-p_v^0$ if it is an ending vertex. For the triangle orientation $(a)$, we can represent such constraints graphically as follows:
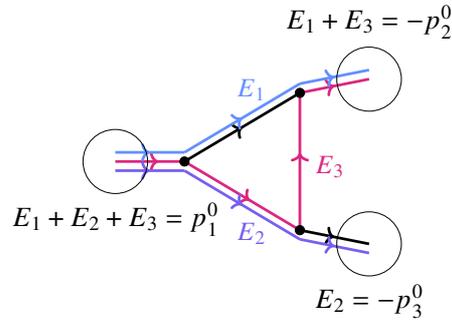
\begin{equation}
\resizebox{6cm}{!}{%
\begin{tikzpicture}
    
    %\node[] (sing) [below left = 2.72cm and 0.5cm of 1] {$ \scalebox{1.3}{\boldsymbol{\otimes}}$};
    %\node[] (3) [below right = 3cm and 3cm of 1] {};
    %\node[] (4) [below left = 3cm and 3cm of 1] {};

    \begin{feynman}
        \vertex(1);
        \vertex[above right = 1.5cm and 2.5cm of 1](2);
        \vertex[below right = 1.5cm and 2.5cm of 1](3);

        \vertex[left = 1.5cm of 1](E1);
        \vertex[above right = 0.3cm and 1.5cm of 2](E2);
        \vertex[below right = 0.3cm and 1.5cm of 3](E3);
        
        \vertex[above=0.2cm of 1] (r11);
        \vertex[above= 0.2cm of 2](r12);
        \vertex[above = 0.2cm of E1](r1E1);
        \vertex[above = 0.2cm of E2](r1E2);
        
        \vertex[below=0.2cm of 1] (r21);
        \vertex[below= 0.2cm of 3](r22);
        \vertex[below = 0.2cm of E1](r2E1);
        \vertex[below = 0.2cm of E3](r2E2);
        
        \vertex[below = 0.75cm of E1](label1) {\LARGE $E_1+E_2+E_3=p_1^0$};
        
        \vertex[above = 0.75cm of E2](label2) {\LARGE$E_1+E_3=-p_2^0$};
        
        \vertex[below = 0.75cm of E3](label2) {\LARGE$E_2=-p_3^0$};

        %\vertex[left = 0.2cm of E1](L1) {\scalebox{1.5}{$p_1$}};
        %\vertex[right = 0.2cm of E2](L2) {\scalebox{1.5}{$p_2$}};
        %\vertex[right = 0.2cm of E3](L3) {\scalebox{1.5}{$p_3$}};
        
        %\vertex[below = 0.3cm of 1](L1_I) {\scalebox{1.5}{$y_1$}};
        %\vertex[above = 0.3cm of 2](L2_I) {\scalebox{1.5}{$y_2$}};
        %\vertex[below = 0.3cm of 3](L3_i) {\scalebox{1.5}{$y_3$}};
        
        \vertex[above right = 1.2cm and 1cm of 1](edge4) {\scalebox{1.5}{$\color{col1}E_1$}};
        \vertex[below right = 1.2cm and 1cm of 1](edge5) {\scalebox{1.5}{$\color{col2}E_2$}};
        \vertex[below right = 1.25cm and 0.2cm of 2](edge6) {\scalebox{1.5}{$\color{col4}E_3$}};

         \diagram*[large]{	
            (1) -- [->-,line width=0.6mm] (2) -- [-<-,line width=0.6mm, col4] (3) -- [-<-,line width=0.6mm, col4] (1),

            (1) -- [-<-,line width=0.6mm, col4] (E1),
            (2) -- [->-,line width=0.6mm, col4] (E2),
            (3) -- [->-,line width=0.6mm] (E3),
            
            (r1E1) -- [->-,line width=0.6mm, col1] (r11) -- [->-,line width=0.6mm, col1] (r12) -- [->-,line width=0.6mm, col1] (r1E2),
            
            (r2E1) -- [->-,line width=0.6mm, col2] (r21) -- [->-,line width=0.6mm, col2] (r22) -- [->-,line width=0.6mm, col2] (r2E2),

        }; 
       \path[draw=black, fill=black] (1) circle[radius=0.1];
       \path[draw=black, fill=black] (2) circle[radius=0.1];
       \path[draw=black, fill=black] (3) circle[radius=0.1];

       \path[draw=black] (E1) circle[radius=0.7];
       \path[draw=black] (E2) circle[radius=0.7];
       \path[draw=black] (E3) circle[radius=0.7];
       
    \end{feynman}

    	%\draw[->, thick, black] (A3) to[out= -10, in=-170] (3);
\end{tikzpicture}
}
\nonumber
\end{equation}
We can parameterize the polytope by 
setting $ \bb E= (E_1,E_2,E_3) = (E,-p_3^0,-p_2^0-E)$ and let $E$ vary between $0$ and $-p_2^0$. The polytope $\mathcal{F}_{G,\bb \sigma}^{\{p^0_a\}}$ is therefore a line segment.  
Using this parameterization, we can explicitly evaluate the Fourier transformation of the flow polytope associated to the digraph above,
\begin{gather}
\widehat{ \mathcal{F}}_{G,\bb \sigma}^{\{p^0_a\}}
(\bb \gamma^\tt   + i \ep \bb 1)
=
\int_{\mathcal{F}_{G,\bb \sigma}^{\{p^0_a\}}}
\dd \bb E\, e^{i \bb E \cdot (\bb \gamma^\tt+ i \ep \bb 1 ) }
=
\int_{0}^{-p_2^0}
\dd E \,e^{i E ( \gamma_1^\tt + i \ep) - ip_3^0 ( \gamma_2^\tt + i \ep) - i(p_2^0+E) (\gamma_3^\tt + i \ep) }
\notag
\\
=
-p_2^0
e^{- ip_3^0 ( \gamma_2^\tt + i \ep) - ip_2^0 (\gamma_2^\tt + \frac12 \gamma_1^\tt  - \frac12 \gamma_3^\tt  + i \ep) }
\sinc\left(
\frac{p_2^0 (\gamma_1^\tt  - \gamma_3^\tt)}{2}
\right)\;,
\label{eq:triangleFourierFlow}
\end{gather}
where $\sinc(x) = \frac{\sin(x)}{x}$. This expression is manifestly bounded as $\sinc(x) \leq 1$. 

Finally, the reduced S-matrix contribution of the digraph above is,
\begin{align}
\label{eq:triangle_px_final}
s_{G,\bb \sigma}(\{p_1\}, \{p_2,p_3\})
=
\int
  \frac{
\left[
\prod_{v\in \{2,3\}} \mathrm{d}^3 \vec{y}_v
\right]
\left[
e^{-i\vec{y}_{2} \cdot \vec{p}_2 -i\vec{y}_{3} \cdot \vec{p}_3}
\right]
 }{8|\vec{z}_4| |\vec{z}_5| |\vec{z}_6|
}
\widehat{\mathcal{F}}_{G,\bb \sigma}^{\{p^0_a\}}(\bb \gamma^\tt + i \ep \bb 1)
\Big|_{\vec y_1 = 0}\;,
\end{align}
where we used the freedom guaranteed by translation invariance to fix one vertex position at the origin, in this case $\vec{y}_1 = 0$.

\subsection{Feynman rules for cut diagrams}
\label{sec:cutrules}
The FOPT Feynman rules for a digraph $(G,\boldsymbol{\sigma})$ with a cut $\mathfrak C$ are:
\begin{enumerate}
\item The integral is $0$ if the closed directed graph $(G,\boldsymbol{\sigma})^\circ$ is not strongly connected or if the admissible paths on the cut do not go from the $\sunny$-side to the $\shady$-side of the graph.
\item Multiply a factor of $-ig$ ($ig$) for each $\sunny$-side ($\shady$-side) interaction vertex. 
\item For each internal vertex $v\in V^{\text{int}}$ of the digraph $(G,\boldsymbol{\sigma})$ integrate over $3$-dimensional space with the measure $2\pi \int \dd^3 \vec{y}_v$.
\item For each edge $e$ of the graph multiply a factor of $\frac{\mp i}{8\pi^2|\vec{z}_e|}$ with a $-$ sign for a $\sunny$-side or a cut edge, and a $+$ sign for a $\shady$-side edge. 
\item For each entirely uncut directed admissible path, $ \pp_\ell$, of $(G,\boldsymbol{\sigma})^\circ$ multiply a factor of 
\begin{align*}
\frac{i}
{
\sum_{e\in \pp_\ell}
|\vec z_e| 
+
\tau_{\mathcal \pp_\ell} 
+i\eta
}
&&\text{ if $\pp_\ell$ consists entirely of $\sunny$-side edges}
\\
\frac{i}
{
-\sum_{e\in \pp_\ell}
|\vec z_e| 
+
\tau_{\pp_\ell} 
+i\eta
}
&&\text{ if $\pp_\ell$ consists entirely of $\shady$-side edges}
\end{align*}
where the sum in the denominator goes over all edges that are in the admissible path $\pp_\ell$ and $\tau_{\mathcal \pp_\ell}$ is the time difference that has passed while going through the $\circ$ vertex, or $0$ if the admissible path does not go through the $\circ$ vertex, i.e.~is a cycle.
\item For each directed admissible path $\pp_{\ell}$ of $(G,\boldsymbol{\sigma})^\circ$ that passes the cut $\mathfrak C$, multiply a factor of 
\begin{align*}
\frac{-2 i |\vec z_{e_\mathfrak C}|}
{
\left(
\sum_{e\in \pp_\ell^\sunny}
|\vec z_e| 
-\sum_{e\in \pp_\ell^\shady}
|\vec z_e| 
+
\tau_{\pp_\ell} 
+i\eta
\right)^2
-
{\vec z_{e_\mathfrak C}}^{\,2}
}\;,
\end{align*}
where we sum over the uncut $\sunny$-side and $\shady$-side edges in $\pp_\ell$, $\pp_\ell^\sunny$ and $\pp_\ell^\shady$, and where $e_\mathfrak C$ denotes the unique edge of the admissible path that is on the cut. 
The edge is unique because, once the path passes over the cut edge, the energy cannot flow back through the cut.
\end{enumerate}

\paragraph{Example} As an example we consider the cut integrals associated to the following graph,
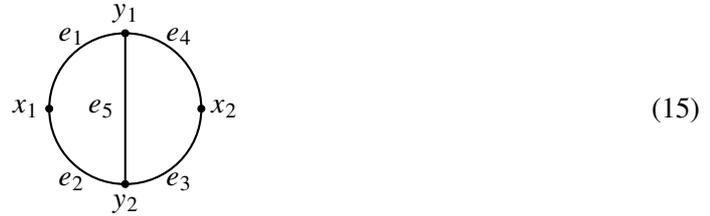
\begin{align}
    \label{eq:theta_graph}
    \def\scale{1}
\begin{tikzpicture}[baseline={([yshift=-0.7ex]0,0)}] 
    \coordinate[label=left:$x_1$] (i1) at (-\scale,0);
    \coordinate[label=right:$x_2$] (i2) at (+\scale,0);
    \coordinate[label=above:$y_1$] (v1) at (0,\scale);
    \coordinate[label=below:$y_2$] (v2) at (0,-\scale);
    \draw[thick] (i1) arc (180:90:\scale) node[midway,above] {$e_1$};
    \draw[thick] (i1) arc (180:270:\scale) node[midway,below] {$e_2$};
    \draw[thick] (i2) arc (0:-90:\scale) node[midway,below] {$e_3$};
    \draw[thick] (i2) arc (0:90:\scale) node[midway,above] {$e_4$};
    \filldraw (v1) circle (1.3pt);
    \filldraw (v2) circle (1.3pt);
    \filldraw (i1) circle (1.3pt);
    \filldraw (i2) circle (1.3pt);
    \draw[thick] (v1) -- (v2) node[midway,left] {$e_5$};
\end{tikzpicture}%
\end{align}
We have the following three different admissible cuts (as permutations of the internal vertices result in topologically indistinguishable graphs),
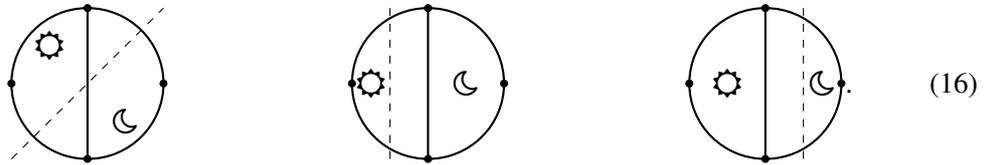
\begin{align}
&    \def\scale{1}
\begin{tikzpicture}[baseline={([yshift=-0.7ex]0,0)}] 
    \coordinate (i1) at (-\scale,0);
    \coordinate (i2) at (+\scale,0);
    \coordinate (v1) at (0,\scale);
    \coordinate (v2) at (0,-\scale);
    \draw[thick] (i1) arc (180:90:\scale);
    \draw[thick] (i1) arc (180:270:\scale);
    \draw[thick] (i2) arc (0:-90:\scale);
    \draw[thick] (i2) arc (0:90:\scale);
    \filldraw (v1) circle (1.3pt);
    \filldraw (v2) circle (1.3pt);
    \filldraw (i1) circle (1.3pt);
    \filldraw (i2) circle (1.3pt);
    \draw[thick] (v1) -- (v2);
    \draw[dashed] (-\scale,-\scale) -- (\scale,\scale);
    \node at (-\scale/2,\scale/2) {$\sunny$};
    \node at (\scale/2,-\scale/2) {$\shady$};
\end{tikzpicture}%
&
&    \def\scale{1}
\begin{tikzpicture}[baseline={([yshift=-0.7ex]0,0)}] 
    \coordinate (i1) at (-\scale,0);
    \coordinate (i2) at (+\scale,0);
    \coordinate (v1) at (0,\scale);
    \coordinate (v2) at (0,-\scale);
    \draw[thick] (i1) arc (180:90:\scale);
    \draw[thick] (i1) arc (180:270:\scale);
    \draw[thick] (i2) arc (0:-90:\scale);
    \draw[thick] (i2) arc (0:90:\scale);
    \filldraw (v1) circle (1.3pt);
    \filldraw (v2) circle (1.3pt);
    \filldraw (i1) circle (1.3pt);
    \filldraw (i2) circle (1.3pt);
    \draw[thick] (v1) -- (v2);
    \draw[dashed] (-\scale/2,-\scale) -- (-\scale/2,\scale);
    \node at (-\scale/4*3,0) {$\sunny$};
    \node at (\scale/2,0) {$\shady$};
\end{tikzpicture}%
&
&    \def\scale{1}
\begin{tikzpicture}[baseline={([yshift=-0.7ex]0,0)}] 
    \coordinate (i1) at (-\scale,0);
    \coordinate (i2) at (+\scale,0);
    \coordinate (v1) at (0,\scale);
    \coordinate (v2) at (0,-\scale);
    \draw[thick] (i1) arc (180:90:\scale);
    \draw[thick] (i1) arc (180:270:\scale);
    \draw[thick] (i2) arc (0:-90:\scale);
    \draw[thick] (i2) arc (0:90:\scale);
    \filldraw (v1) circle (1.3pt);
    \filldraw (v2) circle (1.3pt);
    \filldraw (i1) circle (1.3pt);
    \filldraw (i2) circle (1.3pt);
    \draw[thick] (v1) -- (v2);
    \draw[dashed] (\scale/2,-\scale) -- (\scale/2,\scale);
    \node at (-\scale/2,0) {$\sunny$};
    \node at (\scale/4*3,0) {$\shady$};
\end{tikzpicture}%
.
\intertext{Recall that in addition to the positivity requirements only energy flows from $\sunny$ to $\shady$ are allowed on cut edges. Therefore only the following energy flows are compatible with the cuts and the positive energy requirement:}
\label{eq:theta_cuts1}
&    \def\scale{1}
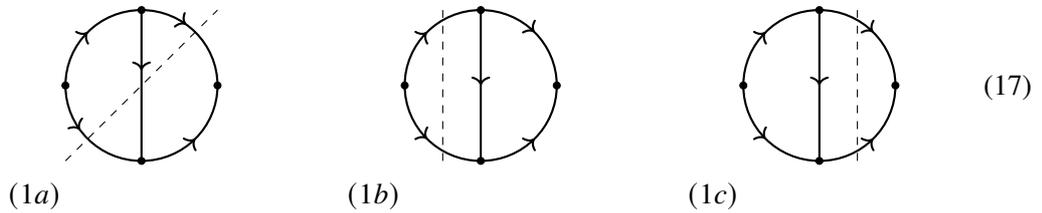
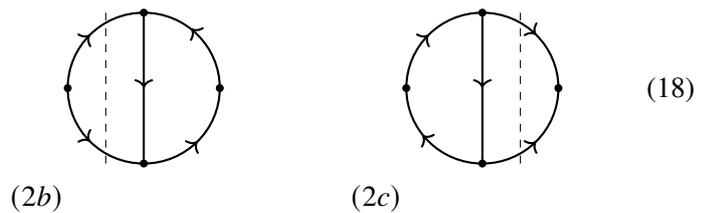
\begin{tikzpicture}[baseline={([yshift=-0.7ex]0,0)}] 
    \coordinate (i1) at (-\scale,0);
    \coordinate (i2) at (+\scale,0);
    \coordinate (v1) at (0,\scale);
    \coordinate (v2) at (0,-\scale);
    \draw[->-,thick] (i1) arc (180:90:\scale);
    \draw[->,thick] (i1) arc (180:270:\scale);
    \draw[->,thick] (v1) arc (90:0:\scale);
    \draw[->-,thick] (v2) arc (-90:0:\scale);
    \draw[->,thick] (v1) -- (v2);
    \filldraw (v1) circle (1.3pt);
    \filldraw (v2) circle (1.3pt);
    \filldraw (i1) circle (1.3pt);
    \filldraw (i2) circle (1.3pt);
    \draw[dashed] (-\scale,-\scale) -- (\scale,\scale);
\end{tikzpicture}%
&
&    \def\scale{1}
\begin{tikzpicture}[baseline={([yshift=-0.7ex]0,0)}] 
    \coordinate (i1) at (-\scale,0);
    \coordinate (i2) at (+\scale,0);
    \coordinate (v1) at (0,\scale);
    \coordinate (v2) at (0,-\scale);
    \draw[->-,thick] (i1) arc (180:90:\scale);
    \draw[->-,thick] (i1) arc (180:270:\scale);
    \draw[->-,thick] (v1) arc (90:0:\scale);
    \draw[->-,thick] (v2) arc (-90:0:\scale);
    \draw[->-,thick] (v1) -- (v2);
    \filldraw (v1) circle (1.3pt);
    \filldraw (v2) circle (1.3pt);
    \filldraw (i1) circle (1.3pt);
    \filldraw (i2) circle (1.3pt);
    \draw[dashed] (-\scale/2,-\scale) -- (-\scale/2,\scale);
\end{tikzpicture}%
&
&    \def\scale{1}
\begin{tikzpicture}[baseline={([yshift=-0.7ex]0,0)}] 
    \coordinate (i1) at (-\scale,0);
    \coordinate (i2) at (+\scale,0);
    \coordinate (v1) at (0,\scale);
    \coordinate (v2) at (0,-\scale);
    \draw[->-,thick] (i1) arc (180:90:\scale);
    \draw[->-,thick] (i1) arc (180:270:\scale);
    \draw[->-,thick] (v1) arc (90:0:\scale);
    \draw[->-,thick] (v2) arc (-90:0:\scale);
    \draw[->-,thick] (v1) -- (v2);
    \filldraw (v1) circle (1.3pt);
    \filldraw (v2) circle (1.3pt);
    \filldraw (i1) circle (1.3pt);
    \filldraw (i2) circle (1.3pt);
    \draw[dashed] (\scale/2,-\scale) -- (\scale/2,\scale);
\end{tikzpicture}%
\\
(1a) && (1b) && (1c)
\notag
\\
\label{eq:theta_cuts2}
&&
&    \def\scale{1}
\begin{tikzpicture}[baseline={([yshift=-0.7ex]0,0)}] 
    \coordinate (i1) at (-\scale,0);
    \coordinate (i2) at (+\scale,0);
    \coordinate (v1) at (0,\scale);
    \coordinate (v2) at (0,-\scale);
    \draw[->-,thick] (i1) arc (180:90:\scale);
    \draw[->-,thick] (i1) arc (180:270:\scale);
    \draw[-<-,thick] (v1) arc (90:0:\scale);
    \draw[->-,thick] (v2) arc (-90:0:\scale);
    \draw[->-,thick] (v1) -- (v2);
    \filldraw (v1) circle (1.3pt);
    \filldraw (v2) circle (1.3pt);
    \filldraw (i1) circle (1.3pt);
    \filldraw (i2) circle (1.3pt);
    \draw[dashed] (-\scale/2,-\scale) -- (-\scale/2,\scale);
\end{tikzpicture}%
&
&    \def\scale{1}
\begin{tikzpicture}[baseline={([yshift=-0.7ex]0,0)}] 
    \coordinate (i1) at (-\scale,0);
    \coordinate (i2) at (+\scale,0);
    \coordinate (v1) at (0,\scale);
    \coordinate (v2) at (0,-\scale);
    \draw[->-,thick] (i1) arc (180:90:\scale);
    \draw[-<-,thick] (i1) arc (180:270:\scale);
    \draw[->-,thick] (v1) arc (90:0:\scale);
    \draw[->-,thick] (v2) arc (-90:0:\scale);
    \draw[->-,thick] (v1) -- (v2);
    \filldraw (v1) circle (1.3pt);
    \filldraw (v2) circle (1.3pt);
    \filldraw (i1) circle (1.3pt);
    \filldraw (i2) circle (1.3pt);
    \draw[dashed] (\scale/2,-\scale) -- (\scale/2,\scale);
\end{tikzpicture}%
\\
&& (2b) && (2c)
\notag
\end{align}
In the picture above, each row features only one orientation of the graph and each column a possible cut. In this example, there are only two admissible paths compatible with a given cut. 
The cut diagram $(1a)$ has the following three routes
{
\def\scale{.8}
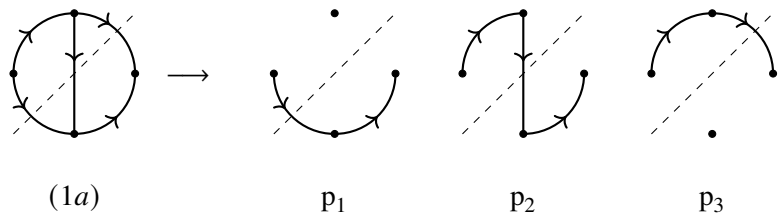
\begin{align*}
\begin{array}{ccccc}
\begin{tikzpicture}[baseline={([yshift=-0.7ex]0,0)}] 
    \coordinate (i1) at (-\scale,0);
    \coordinate (i2) at (+\scale,0);
    \coordinate (v1) at (0,\scale);
    \coordinate (v2) at (0,-\scale);
    \draw[->-,thick] (i1) arc (180:90:\scale);
    \draw[->,thick] (i1) arc (180:270:\scale);
    \draw[->,thick] (v1) arc (90:0:\scale);
    \draw[->-,thick] (v2) arc (-90:0:\scale);
    \draw[->,thick] (v1) -- (v2);
    \filldraw (v1) circle (1.3pt);
    \filldraw (v2) circle (1.3pt);
    \filldraw (i1) circle (1.3pt);
    \filldraw (i2) circle (1.3pt);
    \draw[dashed] (-\scale,-\scale) -- (\scale,\scale);
\end{tikzpicture}%
& \longrightarrow
&\;\;\;\;  \begin{tikzpicture}[baseline={([yshift=-0.7ex]0,0)}] 
    \coordinate (i1) at (-\scale,0);
    \coordinate (i2) at (+\scale,0);
    \coordinate (v1) at (0,\scale);
    \coordinate (v2) at (0,-\scale);
   % \draw[->-,thick] (i1) arc (180:90:\scale);
    \draw[->,thick] (i1) arc (180:270:\scale);
   % \draw[->,thick] (v1) arc (90:0:\scale);
    \draw[->-,thick] (v2) arc (-90:0:\scale);
   % \draw[->,thick] (v1) -- (v2);
    \filldraw (v1) circle (1.3pt);
    \filldraw (v2) circle (1.3pt);
    \filldraw (i1) circle (1.3pt);
    \filldraw (i2) circle (1.3pt);
    \draw[dashed] (-\scale,-\scale) -- (\scale,\scale);
\end{tikzpicture}%
&\;\;\;\;  \begin{tikzpicture}[baseline={([yshift=-0.7ex]0,0)}] 
    \coordinate (i1) at (-\scale,0);
    \coordinate (i2) at (+\scale,0);
    \coordinate (v1) at (0,\scale);
    \coordinate (v2) at (0,-\scale);
    \draw[->-,thick] (i1) arc (180:90:\scale);
   % \draw[->,thick] (i1) arc (180:270:\scale);
 %   \draw[->,thick] (v1) arc (90:0:\scale);
  \draw[->-,thick] (v2) arc (-90:0:\scale);
    \draw[->,thick] (v1) -- (v2);
    \filldraw (v1) circle (1.3pt);
    \filldraw (v2) circle (1.3pt);
    \filldraw (i1) circle (1.3pt);
    \filldraw (i2) circle (1.3pt);
    \draw[dashed] (-\scale,-\scale) -- (\scale,\scale);
\end{tikzpicture}%
&\;\;\;\; \begin{tikzpicture}[baseline={([yshift=-0.7ex]0,0)}] 
    \coordinate (i1) at (-\scale,0);
    \coordinate (i2) at (+\scale,0);
    \coordinate (v1) at (0,\scale);
    \coordinate (v2) at (0,-\scale);
    \draw[->-,thick] (i1) arc (180:90:\scale);
    %\draw[->,thick] (i1) arc (180:270:\scale);
    \draw[->,thick] (v1) arc (90:0:\scale);
   % \draw[->-,thick] (v2) arc (-90:0:\scale);
   % \draw[->,thick] (v1) -- (v2);
    \filldraw (v1) circle (1.3pt);
    \filldraw (v2) circle (1.3pt);
    \filldraw (i1) circle (1.3pt);
    \filldraw (i2) circle (1.3pt);
    \draw[dashed] (-\scale,-\scale) -- (\scale,\scale);
\end{tikzpicture}% 
\\
&&&&
\\
(1a) & &\;\;\;\;\pp_1 &\;\; \;\;\pp_2 &\;\;\;\; \pp_3
\end{array}
\end{align*}
}

Hence, applying the FOPT-cut Feynman rules from above to the cut diagram $(1a)$ 
results in the following expression 
\begin{align}
A_{(\boldsymbol{\sigma},\mathfrak{C})_{(1a)}} 
&=
-8
\frac{(2\pi)^2 g^{4}}{(8\pi^2)^{5}}
\int \frac{\dd^3 \vec y_1 \dd^3 \vec y_2}{|\vec z_1| |\vec z_2| |\vec z_3| |\vec z_4| |\vec z_5| }\times
\\
&\times
\frac{|\vec z_2|}{(-|\vec z_3|+\tau+i\eta)^2-\vec z_2^{\,2}}
\frac{|\vec z_5|}{(|\vec z_1|-|\vec z_3|+\tau+i\eta)^2-\vec z_5^{\,2}}
\frac{|\vec z_4|}{(|\vec z_1|+\tau+i\eta)^2-\vec z_4^{\,2}}\;.
\end{align}
where we accounted for the admissible paths through the cut, $23$, $153$ and $14$ via the appropriate denominators, and $\tau = x_2^0 - x_1^0$. 

One can check that the remaining cut diagrams, which have a different sized cut from $(1a)$, will have the same integral measure as $(1a)$ \cite{Borinsky:2022msp}. This implies that virtual and real IR divergences could cancel locally in FOPT.

\section{Massless fermion lines in FOPT}
In this section we extend the FOPT framework to massless fermion lines. To do so we use that the fermion propagator in coordinate space, $S(x)$, is related to the scalar propagator $\Delta(x)$ by $S(x)=\gamma_\mu \partial^\mu\Delta (x)$. This leads one to modify the intermediate steps of the derivation of FOPT by considering the integral (to be contracted with  $\gamma_\mu$) 
 \begin{align}
 I^\mu_e=\int dz^0_{e}\frac{z_e^\mu\delta\left(z_{e}^0-x^0_e\right)}{(-{z_{e}^0}^2+{\vec{z}_e}^{\;2}+i\eta)^2}=\int_{-\infty}^{+\infty}\frac{dE_e}{2\pi}\int dz^0_{e}\frac{z_e^\mu e^{iE_e(z_{j}^0-x^0_e)}}{(-{z_{e}^0}+{\vec{z}_e}+i\eta)^2({z_{e}^0}+{\vec{z}_e}+i\eta)^2}\;,
\end{align}
for each fermionic edge $e$ of a given diagram, where $x^0_e$ is the time component of the propagator's argument.
We see that the integration in $z_e^0$, after proper closing of the contour of integration, will pick up the residues of the two double poles at $z_e^0=\pm(|\vec{z}_e|+i\eta)$. These are:
\begin{itemize}
\item For spatial components of the numerator (dropping the $i\eta$),  
\begin{equation}
\text{Res}(f,\pm(|\vec{z}_e|+i\eta))=\theta(\pm E_e)z_e^i\Big(\frac{i2|\vec{z}_e|E_e\mp2}{(2|\vec{z}_e|)^3}\Big)e^{\pm iE_e\left(|\vec{z}_e|\mp x^0_e\right)}\;.
\end{equation}
\item For the time component of the numerator, 
\begin{equation}
\text{Res}(f,\pm(|\vec{z}_e|+i\eta))=\theta(\pm E_e)\Big(\frac{\pm iE_e}{4|\vec{z}_e|}\Big)e^{\pm iE_e\left(|\vec{z}_j|\mp x^0_e\right)}\;.
\end{equation}
\end{itemize}
Hence, this integration produces, after defining the lightlike vector $\tilde{z}^{\mu }_{e,\sigma_e=\pm1}=(\pm|\vec{z}_e|,\vec{z}_e)$, 
 \begin{align}
 I^\mu_e&=\frac{i}{(2|\vec{z}_e|)^3}\sum_{\sigma_e=\pm1}\tilde{z}^\mu_{e,\sigma_e}\Big(2|\vec{z}_e|\frac{\partial}{\partial |\vec{z}_e|}-2\sum_{i=1}^3\delta^{\mu i}\Big)\int_{-\infty}^{+\infty}{dE^0_e} \Big({\theta (\sigma_e E^0_e)e^{iE_e^0(\sigma_e|\vec{z}_e|-x^0_e+i\eta)}}\Big) \;,
\end{align}
where  $\sigma_e $ assigns $\pm 1$ to an edge $e$ for a positive or negative energy flow.
This expression can be treated similar to the scalar FOPT case.

Thus, the resulting Feynman rule is that each fermion line, $e$, contributes with an extra factor $$\boxed{\gamma_\mu \frac{{\tilde{z}}_{e,\sigma_e}^\mu}{(2|\vec{z}_e|)^2}\Big(2\sum_{i=1}^3\delta^{\mu i}-2|\vec{z}_e|\frac{\partial}{\partial |\vec{z}_e|}\Big)}\;,$$ where we point out that two upper indices are repeated.

\section{Steps towards FOPT for massive scalar lines in arbitrary dimensions}
Just as to the loop-tree duality \cite{Catani:2008xa,Bierenbaum:2010cy,Capatti:2019ypt,Runkel:2019yrs,Aguilera-Verdugo:2020set,Sborlini:2021owe,Bobadilla:2021pvr,Kromin:2022txz,Berghoff:2022vah}, FOPT has similarities to Light-Cone Ordered Perturbation Theory (LCOPT) \cite{Erdogan:2017gyf}, and most treatments in LCOPT can be extended to FOPT. In \cite{Erdogan:2017gyf}, the inclusion of massive lines and the extension of LCOPT to arbitrary dimensions is performed by using the dispersive representation of a scalar propagator of mass, $m$, in $D=4-2\ep$ dimensions,
\begin{equation}
\Delta(z^2,m)\ =\ \int_0^\infty \frac{dz'{}^2}{\pi }\; \frac{ {\rm Im}\; \Delta \left(z'{}^2+i\eta,m \right)}{-z^2 + z'{}^2+i\eta}\; .
\end{equation}
The imaginary parts for the massless and massive scalar propagators in $D=4-2\ep$ dimensions are given in \cite{Erdogan:2017gyf,Zhang:2008jy}. 
Following this, one must modify eq. (\ref{eq:FI}) as
\begin{align}
A_G(x_1, \ldots, x_{|\gVE|})=
\frac{(-ig)^{|\gVI|}}{(2\pi)^{2|\gE|}}
\left[ \prod_{v \in \gVI}\int \dd^4 y_v \right]
\left[ \prod_{e \in \gE} \int_0^\infty \frac{dz'_e{}^2}{\pi }\frac{{\rm Im}\; \Delta \left(z'_e{}^2+i\eta,m \right)}{-z_e^2+{z_e'}^2+i\eta}\right]\;.\label{eq:FI2}
\end{align}
With this representation, it is possible to perform the full treatment of FOPT to obtain that an orientation, $\bb \sigma$, contributing to a graph $G$ in a massive scalar $D$-dimensional QFT equals
\begin{align}
    A_{G,\bb \sigma}&(x_1, \ldots, x_{|\gVE|})=\nonumber\\
    &=
\frac{(2\pi g)^{|\gVI|}}{(-4\pi^2)^{|\gE|}}
\left( \prod_{v \in \gVI} \int \dd^3 \vec y_v 
\right)
\left(
\prod_{e\in \gE}
 \int_0^\infty \frac{dz'_e{}^2}{\pi }\frac{{\rm Im}\; \Delta \left(z'_e{}^2+i\eta,m \right)}{2 \sqrt{|\vec{z}_e|^2+z_e'{}^2}}
\right)
 \prod_{\pp \in \Gamma} 
\frac{1}
{
\gamma_\pp
+\tau_{\pp} 
+ i \eta
}\;.
\label{eq:delta_free_rep_1}
\end{align}
Where now each path length, $\gamma_\pp$, is modified as
\begin{equation}
    \gamma_\mathrm{p} =\sum_{e\in \pp}
\left(\sqrt{|\vec{z}_e|^2+z_e'{}^2}\right)\;.
\end{equation}

Thus, FOPT can be extended to massive lines and arbitrary dimensions by the inclusion of dispersive integrals and by substituting $|\vec{z}_e|\to\sqrt{|\vec{z}_e|^2+z_e'{}^2}$ for each edge of a given diagram. These dispersive integrals disappear when the massless and four-dimensions limits are taken, since the discontinuity of $\Delta$ vanishes away from the lightcone and approaches a delta function, reproducing the known results of \cite{Borinsky:2022msp}. 
The extensions of FOPT presented in these proceedings are part of ongoing research \cite{Borinsky2}.

\end{document}